# High temperature singlet-based magnetism from Hund's rule correlations

Lin Miao,[1,2] Rourav Basak,[1] Sheng Ran,[3] Yishuai Xu,[1] Erica Kotta,[1] Haowei He,[1] Jonathan D. Denlinger,[2] Yi-De Chuang,[2] Y. Zhao,[3,4] Z. Xu,[3] J. W. Lynn,[3] J. R. Jeffries,[5] S. R. Saha,[3,6] Ioannis Giannakis,[7] Pegor Aynajian,[7] Chang-Jong Kang,[8] Yilin Wang,[9] Gabriel Kotliar,[8] Nicholas P. Butch,[3,6] L. Andrew Wray,[1]*

[1] Department of Physics, New York University, New York, New York 10003, USA
[2] Advanced Light Source, Lawrence Berkeley National Laboratory, Berkeley, CA 94720, USA
[3] NIST Center for Neutron Research, National Institute of Standards and Technology, Gaithersburg, Maryland 20899, USA
[4] Department of Materials Science and Engineering, University of Maryland, College Park, Maryland 20742, USA
[5] Materials Science Division, Lawrence Livermore National Laboratory, Livermore, CA 94550
[6] Center for Nanophysics and Advanced Materials, Department of Physics, University of Maryland, College Park, Maryland 20742, USA
[7] Department of Physics, Applied Physics and Astronomy, Binghamton University, Binghamton, NY 13902
[8] Department of Physics and Astronomy, Rutgers University, Piscataway, New Jersey 08854-8019, USA
[9] Department of Condensed Matter Physics and Materials Science, Brookhaven National Laboratory, Upton, New York 11973, USA

* To whom correspondence should be addressed; E-mail: lawray@nyu.edu.

# Abstract

Uranium compounds can manifest a wide range of fascinating many-body phenomena, and are often thought to be poised at a crossover between localized and itinerant regimes for 5f electrons. The antiferromagnetic dipnictide $USb_2$ has been of recent interest due to the discovery of rich proximate phase diagrams and unusual quantum coherence phenomena. Here, linear-dichroic X-ray absorption and elastic neutron scattering are used to characterize electronic symmetries on uranium in $USb_2$ and isostructural $UBi_2$. Of these two materials, only $USb_2$ is found to enable strong Hund's rule alignment of local magnetic degrees of freedom, and to undergo distinctive changes in local atomic multiplet symmetry across the magnetic phase transition. Theoretical analysis reveals that these and other anomalous properties of the material may be understood by attributing it as the first known high temperature realization of a singlet ground state magnet, in which magnetism occurs through a process that resembles exciton condensation.

# Introduction

Uranium compounds can feature a fascinating interplay of strongly correlated and itinerant electronic physics, setting the stage for emergent phenomena such as quantum criticality, heavy fermion superconductivity, and elusive hidden order states [1-13]. The isostructural uranium dipnictides $UX_2$ (X=As, Sb, Bi) present a compositional series in which high near-neighbor uranium-uranium coordination supports robust planar antiferromagnetism ($T_N$~200K, see Fig. 1a-b) [7,8]. Of these, the $USb_2$ variant has received close attention due to the discovery of several unexplained low temperature quantum coherence phenomena at $T$<100K [7,9,10,11], and a remarkably rich phase diagram incorporating quantum critical and tricritical points as a function of pressure and magnetic field [12,13]. However, the effective valence state of uranium and the resulting crystal field state basis defining the f-electron component of local moment and Kondo physics have not been identified. Here, X-ray absorption (XAS) at the uranium O-edge and numerical modeling are used to evaluate the low energy atomic multiplet physics of $USb_2$ and $UBi_2$, revealing only $USb_2$ to have significant Hund's rule correlations. These investigations yield the prediction that $USb_2$ must be a uniquely robust realization of a singlet-ground-state magnet, in which magnetic moments appear via the occupation of low-energy excited states on a non-magnetic background (see Fig. 1c). The evolution of crystal field symmetries and magnetic ordered moment across the antiferromagnetic phase transition is measured with linear dichroism (XLD) and elastic neutron scattering, confirming that the magnetic transition in $USb_2$ occurs through an exotic process that resembles exciton condensation.

# Results

**Electron configuration on uranium in $UBi_2$ and $USb_2$**

Unlike the case with stronger ligands such as oxygen and chlorine, there is no unambiguously favored effective valence picture for uranium pnictides. Density functional theory suggests that the charge and spin density on uranium are significantly modified by itinerancy effects [14,15] (see also Supplementary Note 1), as we will discuss in the analysis below, making it difficult to address this question from secondary characteristics such as the local or ordered moment. However, analyses in 2014-2016 have shown that resonant fine structure at the O-edge (5d→5f transition) provides a distinctive fingerprint for identifying the nominal valence state and electronic multiplet symmetry on uranium [16-19]. X-ray absorption spectra (XAS) of $UBi_2$ and $USb_2$ were measured by the total electron yield (TEY) method, revealing curves that are superficially similar but quantitatively quite different (Fig. 2a). Both curves have prominent resonance features at $h\nu$~100 and ~113 eV that are easily recognized as the 'R1' and 'R2' resonances split by the G-series Slater integrals [16]. Within models, these resonances are narrowest and most distinct for $5f^0$ systems, and merge as 5f electron number increases, becoming difficult to distinguish beyond $5f^2$ (see Fig. 2a (bottom) simulations). The $USb_2$ sample shows absorption features that closely match the absorption curve of $URu_2Si_2$ [16], and are associated with the $J$=4 ground states of a $5f^2$ multiplet. This correspondence can be drawn with little ambiguity by noting a one-to-one feature correspondence with the fine structure present in a second derivative analysis (SDI, see Fig. 2b).

The R1 and R2 resonances of $UBi_2$ are more broadly separated than in $USb_2$, and the lower energy R1 feature of $UBi_2$ is missing the prominent leading edge peak at $h\nu$~98.2 eV (peak-B), which is a characteristic feature of $5f^2$ uranium [16, 17]. The $UBi_2$ spectrum shows relatively little intensity between R1 and R2, and the higher energy R2 resonance has a much sharper intensity onset. All of these features are closely consistent with expectations for a $5f^1$ multiplet, and the SDI curve in Fig. 2b reveals that the R1 fine structure of $UBi_2$ is a one-to-one match for the $5f^1$ multiplet. We note that a close analysis is not performed for R2 as it is influenced by strong Fano interference (see Supplementary Note 2). The lack of prominent $5f^2$ multiplet features suggests that the $5f^1$ multiplet state is quite pure, and the measurement penetration depth of several nanometers (see Methods) makes it unlikely that this distinction between UHV-cleaved $UBi_2$ and $USb_2$ originates from surface effects. However, the picture for $UBi_2$ is complicated by a very rough cleaved surface, which our STM measurments (see Supplementary Note 3) find to incorporate at least two non-parallel cleavage planes. Surface oxidation in similar compounds is generally associated with the formation of $UO_2$ ($5f^2$) and does not directly explain the observation of a $5f^1$ state.

We note that even with a clean attribution of multiplet symmetries, it is not at all clear how different the f-orbital occupancy will be for these materials, or what magnetic moment should be expected when the single-site multiplet picture is modified by band-structure-like itinerancy [10,11] (see also Supplementary Note 1). The effective multiplet states identified by shallow-core-level spectroscopy represent the coherent multiplet (or angular moment) state on the scattering site and its surrounding ligands, but are relatively insensitive to the degree of charge transfer from the ligands [20].

Nonetheless, the $5f^1$ and $5f^2$ nominal valence scenarios have very different physical implications. A $5f^1$ nominal valence state does not incorporate multi-electron Hund's rule physics [21,22] (same-site multi-electron spin alignment), and must be magnetically polarizable with non-zero pseudospin in the paramagnetic state due to Kramer's degeneracy (pseudospin ½ for the $UBi_2$ crystal structure). By contrast in the $5f^2$ case one expects to have a Hund's metal with strong alignment of the 2-electron moment (see dynamical mean field theory (DFT+DMFT) simulation below), and the relatively low symmetry of the 9-fold ligand coordination around uranium strongly favors a *non-magnetic singlet crystal electric field (CEF) ground state* with $\Gamma_1$ symmetry, gapped from other CEF states by roughly 1/3 the total spread of state energies in the CEF basis (see Table 1). The $\Gamma_1$ state contains equal components of diametrically opposed large-moment $|m_J=+4\rangle$ and $|m_J=-4\rangle$ states, and is poised with no net moment by the combination of spin-orbit and CEF interactions. This unusual scenario in which magnetic phenomena emerge in spite of a non-magnetic singlet ground state has been considered in the context of mean-field models [23-26], and appears to be realized at quite low temperatures (typically $T<\sim10K$) in a handful of rare earth compounds. The resulting magnetic phases are achieved by partially occupying low-lying magnetic excited states, and have been characterized as spin exciton condensates [23].

**Multiplet symmetry from XLD versus temperature**

To address the role of low-lying spin excitations, it is useful to investigate the interplay between magnetism and the occupied multiplet symmetries by measuring the polarization-resolved XAS spectrum as a function of temperature beneath the magnetic transition. Measurements were performed with linear polarization set to horizontal (LH, near z-axis) and vertical (LV, a-b plane) configurations. In the case of UBi$_2$, the XAS spectrum shows little change as a function of temperature from 15 to 210K (see Fig. 3a-b), and temperature dependence in the dichroic difference (XLD, see Fig. 3b) between these linear polarizations is inconclusive, being dominated by noise from the data normalization process (see Methods and Supplementary Note 4). This lack of temperature dependent XLD is consistent with conventional magnetism from a doublet ground state. The XLD matrix elements do not distinguish between the up- and down-moment states of a Kramers doublet, and so strong XLD is only expected if the magnetic phase incorporates higher energy multiplet symmetries associated with excitations in the paramagnetic state.

By contrast, the temperature dependence of USb$_2$ shows a large monotonic progression (Fig. 3c-d), suggesting that the atomic symmetry changes significantly in the magnetic phase. The primary absorption peak ($h\nu$~98.2eV, peak-B) is more pronounced under the LH-polarization at low-temperature, and gradually flattens as temperature increases. The LV polarized spectrum shows the opposite trend, with a sharper peak-B feature visible at high temperature, and a less leading edge intensity at low temperature. This contrasting trend is visible in the temperature dependent XLD in Fig. 3d, as is a monotonic progression with the opposite sign at peak-C ($h\nu$~100.8eV).

Augmenting the atomic multiplet model for $5f^2$ uranium with mean-field magnetic exchange (AM+MF) aligned to match the $T_N$~203K phase transition (see Methods) results in the temperature dependent XAS trends shown in Fig. 3e. The temperature dependent changes in peak-B and peak-C in each linear dichroic curve match the sign of the trends seen in the experimental data, but occur with roughly twice the amplitude, as can be seen in Fig. 3d,f. No attempt is made to precisely match the $T$>200K linear dichroism, as this is influenced by itinerant and Fano physics not considered in the model. The theoretical amplitude could easily be reduced by adding greater broadening on the energy loss axis or by fine tuning of the model (which has been avoided – see Methods). However, it is difficult to compensate for a factor of two, and the discrepancy is likely to represent a fundamental limitation of the non-itinerant mean field atomic multiplet model. Indeed, when the competition between local moment physics and electronic itinerancy is evaluated for USb$_2$ with dynamical mean field theory (DFT+DMFT), we find that the uranium site shows a non-negligible ~25% admixture of $5f^1$ and $5f^3$ configurations (see Fig. 4a).

**Magnetic ordered moment and the nature of fluctuations**

Compared with conventional magnetism, the singlet ground state provides a far richer environment for low temperature physics within the magnetic phase. In a conventional magnetic system, the energy gap between the ground state and next excited state grows monotonically as temperature is decreased beneath the transition, giving an increasingly inert many-body environment. However, in the case of singlet ground state magnetism, the ground state is difficult to magnetically polarize, causing *the energy gap between the ground state and easily polarized excited states to shrink* as temperature is lowered and the magnetic order parameter becomes stronger. Consequently, within the AM+MF model, many states keep significant partial occupancy down to $T<100K$, and the first excited state (derived from the $\Gamma_5$ doublet) actually grows in partial occupancy beneath the phase transition (see Fig. 4b). Of the low energy CEF symmetries (tracked in Fig. 4b), $\Gamma_5$ and $\Gamma_2$ are of particular importance, as $\Gamma_5$ is a magnetically polarizable Ising doublet, and $\Gamma_2$ is a singlet state that can partner coherently with the $\Gamma_1$ ground state to yield a z-axis magnetic moment (see Supplementary note 5). These non-ground-state crystal field symmetries retain a roughly $1/3^{rd}$ of the total occupancy at $T=100K$, suggesting that a heat capacity peak similar to a Schottky anomaly should appear at low temperature, as has been observed at $T<\sim50K$ in experiments (see supplement of Ref. [10]). Alternatively, when intersite exchange effects are factored in, the shrinking energy gap between the $\Gamma_1$ and $\Gamma_5$ CEF states at low temperature will enable Kondo-like resonance physics and coherent exchange effects that are forbidden in conventional magnets.

Critical behavior at the Néel transition should also differ, as the phase transition in a singlet-ground-state magnet is only possible on a background of strong fluctuations. Measuring the ordered moment as a function of temperature with elastic neutron scattering (see Fig. 4c) reveals that the $UBi_2$ moment follows a trend that appears consistent with the $\beta=0.327$ critical exponent for a 3D Ising system [27]. The order parameter in $USb_2$ has a sharper onset that cannot be fitted sufficiently close to the transition point due to disorder, but can be overlaid with an exponent of $\beta \sim 0.19$, and may resemble high-fluctuation scenarios such as tricriticality ($\beta=0.25$ [28, 29]). This sharp onset cannot be explained from the AM+MF model (blue curve in Fig. 4c), as mean field models that replace fluctuations with a static field give large critical exponents such as $\beta=0.5$ and unphysically high transition temperatures in systems where fluctuations are important. Another approach to evaluate the importance of fluctuations is to lower the Néel temperature by alloying with non-magnetic thorium (Th), as $U_{1-x}Th_xSb_2$ (see Fig. 4d), thus quenching thermal fluctuations at the phase transition. Performing such a growth series reveals that the magnetic transition can be suppressed to $T_N \sim 100K$, but is then abruptly lost at $x \sim 0.7$, consistent with the need for fluctuations across a CEF gap of $k_B T_N \sim 10$ meV, which matches expectations from theory for the energy separation between $\Gamma_1$ and $\Gamma_5$ (see Table 1 and Methods).

## Discussion

In summary, we have shown that the $USb_2$ and $UBi_2$ O-edge XAS spectra represent different nominal valence symmetries, with $USb_2$ manifesting $5f^2$ moments that are expected to create a Hund's metal physical scenario, and $UBi_2$ showing strong $5f^1$ –like symmetry character. The CEF ground state of a paramagnetic $USb_2$ Hund's metal is theoretically predicted to be a robust non-magnetic singlet, creating an exotic setting for magnetism that resembles an exciton condensate, and is previously only known from fragile and low temperature realizations. The temperature dependence of XLD measurements is found to reveal a symmetry evolution consistent with singlet-based magnetism. Neutron diffraction measurements show a relatively sharp local moment onset at the transition, consistent with the importance of fluctuations to nucleate the singlet-based magnetic transition, and suppressing thermal fluctuations in a doping series is found to quench magnetism beneath $T_N < \sim 100K$.

Taken together, these measurements are consistent with a singlet-based magnetic energy hierarchy that yields an anomalously large number of thermally accessible degrees of freedom at low temperature (*T*<100K), and provides a foundation for explaining the otherwise mysterious coherence effects found in previous transport, heat capacity, and ARPES measurements at *T*<100K [7,9,10,11]. The interchangeability of elements on both the uranium (demonstrated as $U_{1-x}Th_xSb_2$) and pnictogen site suggests $UX_2$ as a model system for exploring the crossover into both Hund's metal and singlet-ground-state magnetic regimes.

## Methods

**Experiment**

The samples of $UBi_2$ and $USb_2$ were top-posted in a nitrogen glove-box and then transferred within minutes to the ultra high vacuum (UHV) environment. The samples were cleaved in UHV and measured *in-situ*, with initial U O-edge spectra roughly 30 minutes after cleavage. The UV-XAS measurements were performed in the MERIXS (BL4.0.3) in the Advanced Light Source with base pressure better than $4\times10^{-10}$ Torr. The switch between linear horizontal polarization (LH-pol) and linear vertical polarization (LV-pol) is controlled by an elliptically polarizing undulator (EPU) and keep precisely the same beam spot before and after the switch. The incident angle of the photon beam was 30°, which gives a 75% out-of-plane E-vector spectral component under the LH-pol condition and 100% in-plane E-vector under the LV-pol condition. The XAS signal was collected by the total electron yield (TEY) method. The penetration depth of VUV and soft X-ray XAS measured with the TEY method is generally in the 2-4 nm range set by the mean free path of low energy ($E$<~10eV) secondary electrons created in the scattering process [30], making it a much more bulk sensitive technique than single-particle techniques such as angle resolved photoemission.

Air-exposed $UBi_2$ can degrade rapidly due to oxidization. No evidence of a large volume fraction of oxide or other phases was found from neutron scattering data for $USb_2$ and $UBi_2$. Possible sample oxidation was surveyed by measuring oxygen $L_1$-edge XAS via TEY for both $USb_2$ and $UBi_2$ during the uranium O-edge XAS experiments. An oxygen $L_1$-edge signal was visible at the cleaved surface of both samples, and found to have similar intensity for both

USb$_2$ and UBi$_2$ samples (Supplementary note 6).

The O-edge XAS curves observed under LH-pol and LV-pol polarization are normalized by assigning constant intensity to the integrated area of the R1 region. Spectral intensity was integrated between featureless start (95eV) and end points (102eV) for both UBi$_2$ and USb$_2$. The linear dichroism of the XAS in the main text is defined as:

$$I_{LD} = (I_{LH} - I_{LV})/I_{LH(max)} \quad (1)$$

whereas $I_{LH(max)}$ is the XAS intensity maximum under LH-pol condition within R1 region. The monotonic temperature linear dichroism of USb$_2$ in the main text is a solid result under different data normalization process but linear dichroic rate can be influenced by some factors, for example the irreducible background in $I_{LH(max)}$. In the simulation, the tuning the broadening factor is also easy to change simulated linear dichroic rate which make seriously quantitative comparison of the linear dichroism between experiment and the simulation meaningless.

Neutron diffraction measurements were performed on single crystals at the BT-7 thermal triple axis spectrometer at the NIST Center for Neutron Research [31] using a 14.7meV energy and collimation: open - 25' - sample - 25' - 120'. For USb$_2$, the magnetic intensity at the (1, 0, 0.5) peak was compared to the nuclear intensity at the (1, 0, 1) peak, while the temperature dependence of the (1 1, 0.5) peak was used to calculate an order parameter. For UBi$_2$, the temperature-dependent magnetic intensity at the (1, 1, 1) peak was compared to nuclear intensity at (1, 1, 1) peak at 200K, above the Néel temperature. In both cases, an $f^2$ magnetic form factor was assumed [32].

**Atomic Multiplet + Mean Field Model (AM+MF)**

Atomic multiplet calculations were performed as in ref. [16], describing $5d^{10}5f^n \rightarrow 5d^9 5f^{n+1}$ X-ray absorption in the dipole approximation. Hartree-Fock parameters were obtained from the Cowan code [33], and full diagonalization of the multiplet Hamiltonian was performed using LAPACK drivers [34]. Hartree-Fock parameters for 5$f$ multipole interactions renormalized by a factor of $\beta$=0.7 for UBi$_2$, and a more significant renormalization of $\beta$=0.55 was found to improve correspondence for USb$_2$. This difference matches the expected trend across a transition between $5f^2$ and $5f^1$ local multiplet states. Core-valence multipole interactions renormalized by $\beta_C$=0.55, consistent with other shallow core hole actinide studies [35]. The 5$f$ spin orbit is not renormalized in USb$_2$ but renormalized by a factor of 1.15 in UBi$_2$ due to the much larger spin orbit coupling on bismuth. A detailed comparison of simulation results generated from two sets of Hartree-Fock parameters is included in Supplementary Note 7.

Total electron yield is dominated by secondary electrons following Auger decay of the primary scattering site. We have assigned core hole lifetime parameters to describe this decay, and adopted the common approximation that the number of secondary electrons escaping from the material following each core hole decay event is independent of the incident photon energy. For the $5f^1$ simulation, the core hole inverse lifetime is $\Gamma$=1.4 eV at $h\nu$<100eV, $\Gamma$=1.8 eV at 100 eV<$h\nu$<108.5 eV, and 6.5 eV at $h\nu$>108.5 eV. For $5f^2$ and $5f^3$ simulations, feature widths were obtained from a core hole inverse lifetime set to $\Gamma$=1.3eV ($h\nu$<99eV), $\Gamma$=1.5eV (99 eV<$h\nu$<103.5 eV), and 6.5 eV ($h\nu$>103.5 eV). In the $5f^3$ simulation, assigning the 103 eV XAS feature to R1 (longer lifetime) as in the Fig. 2 makes it more prominent than if it is assigned to R2 (shorter lifetime). It is also worth noting that scenarios intermediate to $5f^2$ and $5f^3$ do not necessarily closely resemble the $5f^3$ endpoint, and spectral weight in the 103 eV $5f^3$ XAS peak may depend significantly on local hybridization. However, in real materials, $5f^3$ character is associated with a downward shift in the R1 resonance onset energy that is opposite to what is observed in our data [36].

The mean field model was implemented by considering the USb$_2$ uranium sublattice with Ising exchange coupling between nearest neighbors:

$$\mathbf{H}= \sum_i \mathbf{A}_{,i} + \sum_{<i,k>} \mathbf{J}_{i,k}\, \mathbf{S}_{z,i}\, \mathbf{S}_{z,k} \qquad (2)$$

where $\mathbf{A}_{,i}$ is the $5f^2$ single-atom multiplet Hamiltonian, $\mathbf{J}_{i,k}$ is an exchange coupling parameter with distinct values for in-plane versus out-of-plane nearest neighbors, and $\mathbf{S}_{z,i}$ is the z-moment spin operator acting on site $i$. Mean field theory allows us to replace one of the spin interaction terms ($\mathbf{S}_{z,k}$) with a temperature-dependent expectation value, and describe the properties of the system in terms of a thermally weighted single-atom multiplet state ensemble. The specific values of individual $J_{i,k}$ terms are unimportant in this approximation, however their signs must match the antiferromagnetic structure in Fig. 1, and the sum of the absolute value of near-neighbor terms must equal $\mathbf{J}_{eff}=\sum_{<k>} |J_{n,k}|$ = 43 meV to yield a magnetic transition at $T_N$=203K. When considering the doped case of $U_{1-x}Th_xSb_2$, the expectation value $<S_{z,k}>$ is effectively reduced by weighting in the appropriate density of 0-moment $5f^0$ Th sites.

The CEF energy hierarchy has not been fine tuned. Perturbation strengths are scaled to set the lowest energy excitation to 10 meV, a round number that roughly matches the lowest $K_BT_N$ value at which a magnetic transition is observed in $U_{1-x}Th_xSb_2$. This assignment gives a total energy scale for crystal field physics that is approximately comparable to room temperature ($\Delta CEF \sim k_B T_N$), as expected for this class of materials, and the associated orbital energies were found to correspond reasonably (within $<\sim 30\%$) with coarse estimates from density functional theory. The crystal field parameters are listed in the first column of Table 1.

The low temperature ordered moment of M=1.90 $\mu_B$ seen by neutron scattering is matched by downward-renormalizing the moment calculated in the mean field model to 62% (see Fig. 4d shading). Within density functional theory (DFT) models, the consideration of itinerant electronic states provides a mechanism to explain most of this discrepancy. In DFT simulations, the spin component of the magnetic moment is enhanced to $M_S \sim 2$ $\mu_B$ [14-15], larger than the maximal value of $M_S \sim 1.4$ $\mu_B$ that we find in the $5f^2$ (J=4) atomic multiplet picture. This larger DFT spin moment is directly opposed to the orbital magnetic moment, resulting in a smaller overall ordered moment. The ordered moment in the multiplet simulation could alternatively be reduced by strengthening the crystal field, but this is challenging to physically motivate, and has the opposite effect of reducing the spin moment to $M_S < 1$ $\mu_B$.

**Density functional theory + dynamical mean field theory (DFT+DMFT)**

The combination of density functional theory (DFT) and dynamical mean-field theory (DMFT) [37], as implemented in the full-potential linearized augmented plane-wave method [38,39], was used to describe the competition between the localized and itinerant nature of

5*f*-electron systems. The correlated uranium 5*f* electrons were treated dynamically by the DMFT local self-energy, while all other delocalized spd electrons were treated on the DFT level. The vertex corrected one-crossing approximation [38] was adopted as the impurity solver, in which full atomic interaction matrix was taken into account. The Coulomb interaction *U* = 4.0 eV and the Hund's coupling *J* = 0.57 eV were used for the DFT+DMFT calculations.

**Acknowledgements:**

We are grateful for discussions with S. Roy and L. Klein. This research used resources of the Advanced Light Source, which is a DOE Office of Science User Facility under contract no. DE-AC02-05CH11231. Work at NYU was supported by the MRSEC Program of the National Science Foundation under Award Number DMR-1420073. P.A. acknowledges funding from the U.S. National Science Foundation CAREER under award No. NSF-DMR 1654482. The identification of any commercial product or trade name does not imply endorsement or recommendation by the National Institute of Standards and Technology. G.K. and C.-J.K. are supported by DOE BES under Grant No. DE-FG02-99ER45761. G.K. carried out this work during his sabbatical leave at the NYU Center for Quantum Phenomena, and gratefully acknowledges NYU and the Simons foundation for sabbatical support. Y.W. was supported by the US Department of energy, Office of Science, Basic Energy Sciences as a part of the Computational Materials Science Program through the Center for Computational Design of Functional Strongly Correlated Materials and Theoretical Spectroscopy.

**Author Contributions:**

L.M., R.B., Y.X., E.K. and H.H. carried out the XAS experiments with support from J.D., Y.-D.C. and J.R.J.; neutron measurements were performed by S.R. and N.P.B. with support from Y.Z., Z.X., and J.W.L.; STM measurements were performed by I.G., with guidance from P.A.; high quality samples were synthesized by S.R. and S.R.S. with guidance from N.B.; multiplet simulations were performed by L.M. with guidance from L.A.W., and DFT+DMFT simulations were performed by C.-J.K. with assistance from Y.W. and guidance from G.K.; L.M., R.B., Y.X., P.A., N.P.B., and L.A.W participated in the analysis, figure planning and draft preparation; L.A.W. was responsible for the conception and the overall direction, planning and integration among different research units.

**Figure 1**

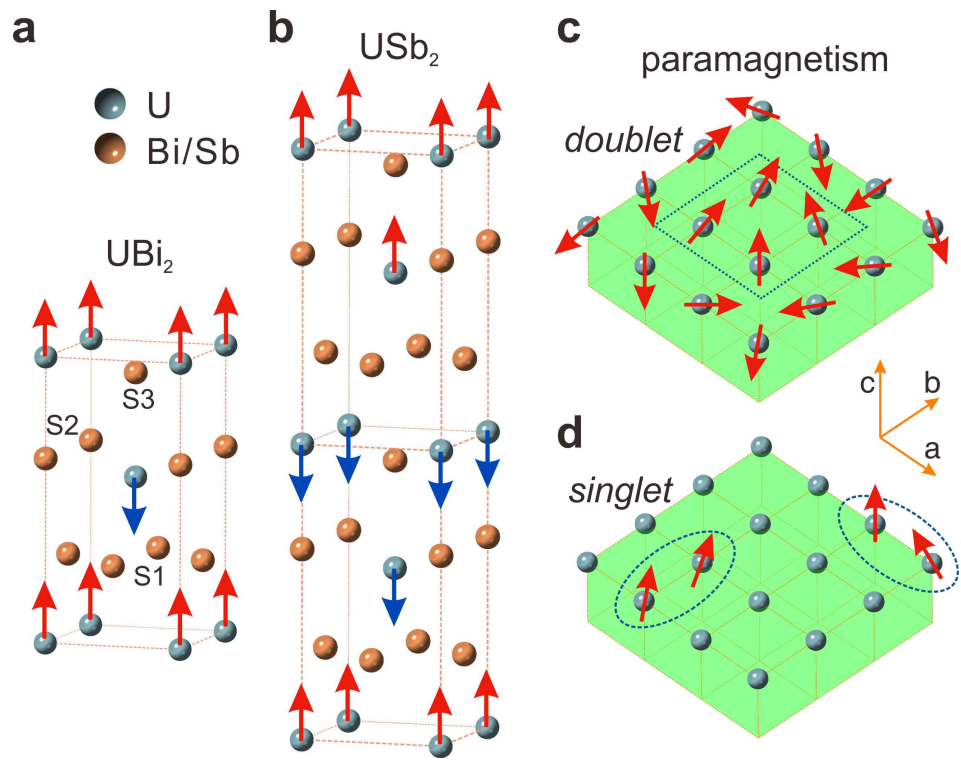

**Figure 1: Singlet ground state magnetism and the ligand cage of U(Bi/Sb)$_2$. a-b,** The U(Sb/Bi)$_2$ crystal structure is shown with spins indicating the antiferromagnetic structure in UBi$_2$ ($T_N$~180K) and USb$_2$ ($T_N$~203K). The uranium atoms have 9-fold ligand coordination with base (S1), middle (S2) and pinnacle (S3) ligand layers as labeled in panel (a) with respect to the central uranium atom. **c-d**, In-plane ferromagnetic nucleation regions are circled in (c) doublet and (d) singlet ground state magnetic systems. The singlet crystal field ground state has no local moment, causing much of the lattice to have little or no magnetic polarization.

**Figure 2**

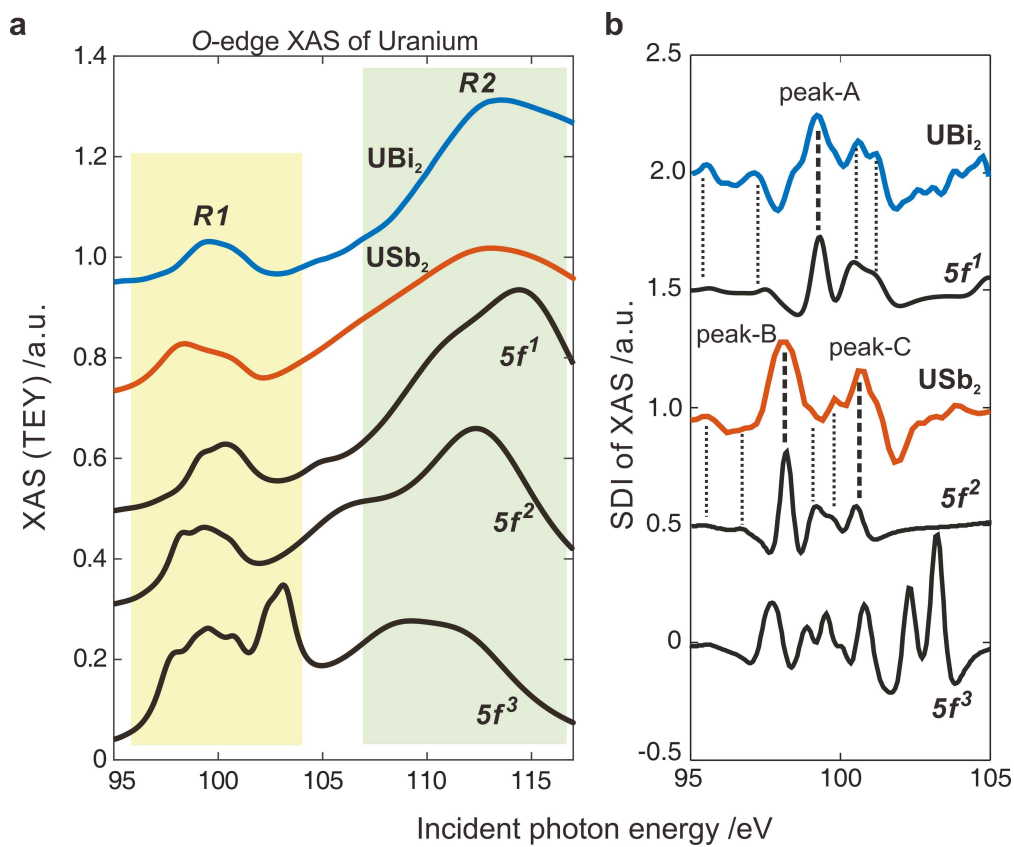

**Figure 2: XAS fine structure and valence of UBi$_2$ and USb$_2$. a**, The x-ray absorption of UBi$_2$ and USb$_2$ on the O-edge of uranium is compared with (bottom) multiplets simulations for $5f^1$ (U$^{5+}$), $5f^2$ (U$^{4+}$) and $5f^3$ (U$^{3+}$). **b**, A negative second derivative (SDI) of the XAS data and simulated curves, with drop-lines showing feature correspondence. Noise in the SDI has an amplitude comparable to the plotted line thickness, and all features identified with drop-lines were consistently reproducible when moving the beam spot. Prominent absorption features are labeled peak-A (UBi$_2$, $h\nu$=99.2eV), peak-B (USb$_2$, $h\nu$=98.2eV) and peak-C (USb$_2$, $h\nu$=100.8eV). Source data are provided as a Source Data file.

**Figure 3**

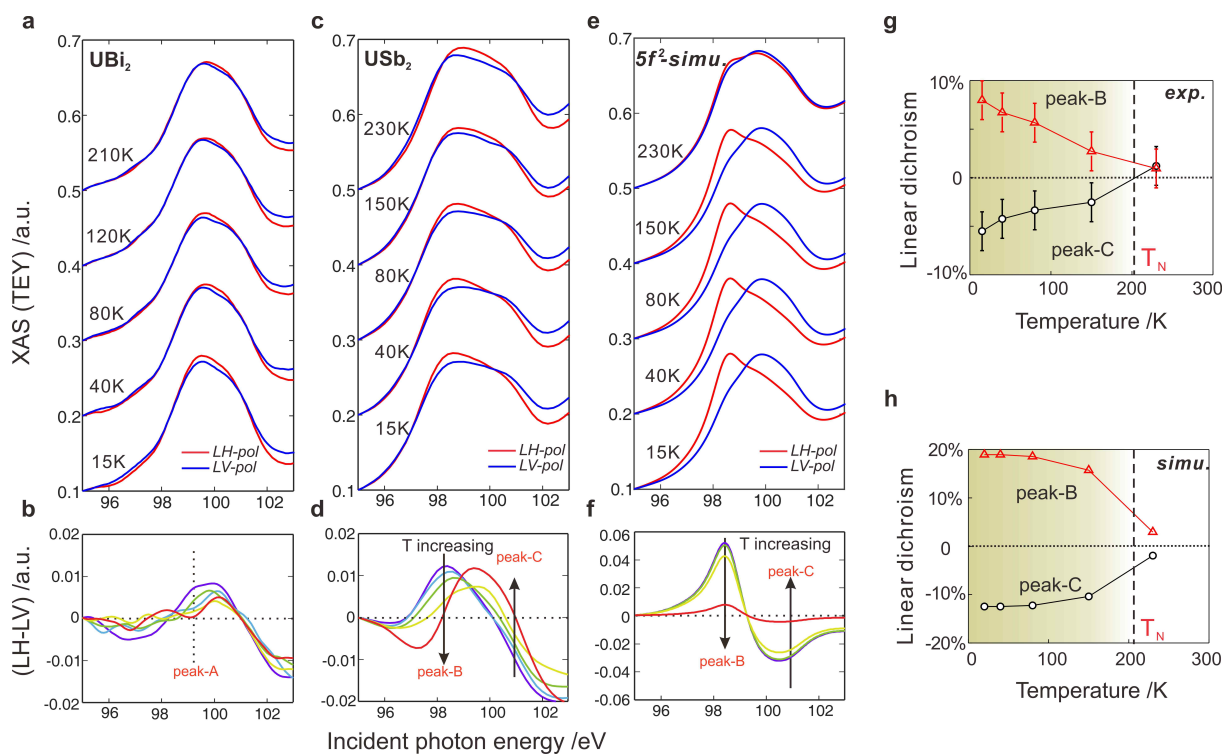

**Figure 3: Temperature dependence of occupied f-electron symmetries. a**, The R1 XAS spectrum of $UBi_2$ is shown for linear horizontal (LH) and vertical (LV) polarizations. **b**, The dichroic difference (LH-LV) is shown with temperature distinguished by a rainbow color order (15K (purple), 40K (blue), 80K (green), 120K (yellow) and 210K (red)). **c-d**, Analogous spectra are shown for $USb_2$. Arrows in panel (d) show the monotonic trend direction on the peak-B and peak-C resonances as temperature increases. **e-f**, Simulations for $5f^2$ with mean-field magnetic interactions. **g**, A summary of the linear dichroic difference on the primary XAS resonances of $USb_2$, as a percentage of total XAS intensity at the indicated resonance energy (hv=98.2eV for peak-B, and hv=100.8eV for peak-C). Error bars represent a rough upper bound on the error introduced by curve normalization. **h**, The linear dichroic difference trends from the mean field model. Source data are provided as a Source Data file. Shading in (g-h) indicates the onset of a magnetic ordered moment.

Figure 4

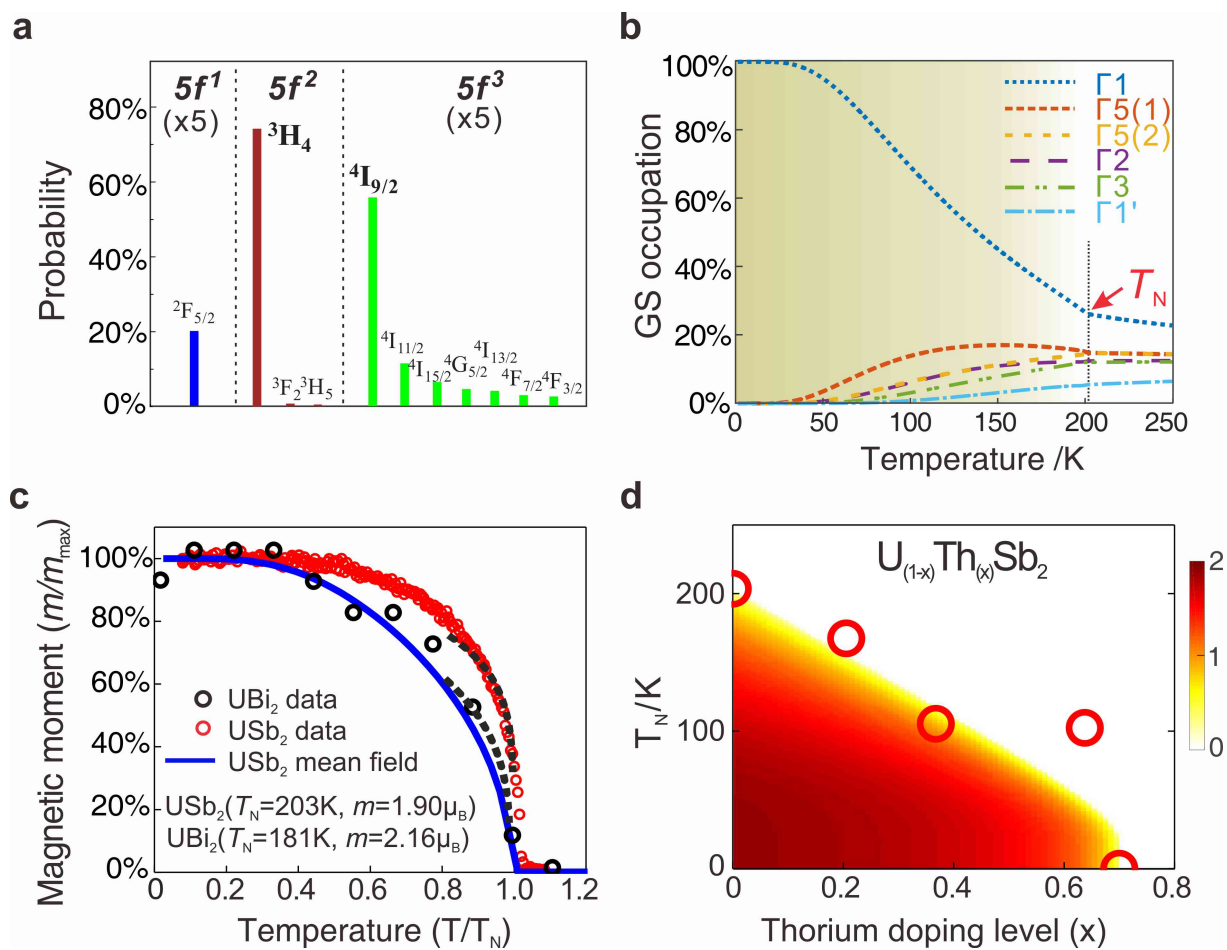

**Figure 4: Electronic symmetry convergence in USb$_2$. a**, The partial multiplet state occupancy on uranium in USb$_2$ from DFT+DMFT numerics, with Hund-aligned symmetries highlighted in bold. **b**, Temperature dependence of the partial occupancy of different multiplet states within a 5$f^2$ mean field model. In spite of a magnetic transition above 200K, roughly 1/3$^{rd}$ of the ground state convergence occurs in the range from 30-100K. The labeled CEF symmetries are only fully accurate in the high temperature paramagnetic state. Beneath the Néel temperature, the Γ$_1$ ground state is magnetically polarized by admixture with Γ$_2$. Shading indicates the onset of a magnetic ordered moment. **c**, The ordered magnetic moment of (red circles) USb$_2$ and (black circles) UBi$_2$ from elastic neutron scattering. The mean field multiplet model for USb$_2$ is shown as a solid blue curve, and critical exponent trends near the phase transition are traced with dashed black lines representing $m(T)=m_{max}(1-T/T_N)^\beta$. The USb$_2$ data are overlaid with a steep critical exponent trend of $\beta$=0.19 indicating strong fluctuations, and the UBi$_2$ data are overlaid with the conventional 3D Ising critical exponent ($\beta$=0.327). **d**, The Néel temperature as a function of doping level in U$_{(1-x)}$Th$_{(x)}$Sb$_2$ (red circles), and the simulated ordered moment in Bohr magnetons (renormalized to 62% as described in Methods; red-hot shading). Source data for all curves are provided as a Source Data file.

**Table 1**

|  | CEF(1) (20/33/33) | CEF(2) (33/33/33) | CEF(3) (50/33/33) | CEF(4) (80/130/130) |
|---|---|---|---|---|
| $\Gamma_1(1)$ | 0 | 0 | 0 | 0 |
| $\Gamma_5(2)$ | 10.0 | 11.4 | 12.6 | 38.5 |
| $\Gamma_2(1)$ | 13.1 | 10.8 | 8.8 | 51.6 |
| $\Gamma_3(1)$ | 13.6 | 13.9 | 15.2 | 54.3 |
| $\Delta$CEF | 27.2 | 30.8 | 37.5 | 106.0 |

**Table 1. The CEF energy hierarchy in USb$_2$.** The energies in millielectron volts of low-lying $5f^2$ multiplet symmetries are shown for four crystal field parameter sets. Parameters in the first column (CEF(1)) follow the relative energy ordering suggested in Ref. [8] (S1<S2~S3, as the S1 bond is relatively short), and are used for all simulations. The state symmetries are summarized in Supplementary Note 5, which includes an energy level diagram. $\Delta$CEF is defined as the gap between the highest energy $J$=4 CEF state and the ground state. CEF parameters listed as (S1/S2/S3) for the sites defined in Fig. 1a. These values have units of millielectron volts, and define delta function potentials for Sb atoms in the (S1) base, (S2) middle, and (S3) c-axis pinnacle of the Sb$_9$ cage around each uranium atom. Specifically, the energy parameters indicate the energy added by a single Sb atom to an $m_j$=0 f-orbital oriented along the U-Sb axis. Source data are provided as a Source Data file.

*Supplementary information for*

# High temperature singlet-based magnetism from Hund's rule correlations

Lin Miao,[1,2] Rourav Basak,[1] Sheng Ran,[3] Yishuai Xu,[1] Erica Kotta,[1] Haowei He,[1] Jonathan D. Denlinger,[2] Yi-De Chuang,[2] Y. Zhao,[3,4] Z. Xu,[3] J. W. Lynn,[3] J. R. Jeffries,[5] S. R. Saha,[3,6] Ioannis Giannakis,[7] Pegor Aynajian,[7] Chang-Jong Kang,[8] Yilin Wang,[9] Gabriel Kotliar,[8] Nicholas P. Butch,[3,6] L. Andrew Wray,[1]*

[1] Department of Physics, New York University, New York, New York 10003, USA
[2] Advanced Light Source, Lawrence Berkeley National Laboratory, Berkeley, CA 94720, USA
[3] NIST Center for Neutron Research, National Institute of Standards and Technology, Gaithersburg, Maryland 20899, USA
[4] Department of Materials Science and Engineering, University of Maryland, College Park, Maryland 20742, USA
[5] Materials Science Division, Lawrence Livermore National Laboratory, Livermore, CA 94550
[6] Center for Nanophysics and Advanced Materials, Department of Physics, University of Maryland, College Park, Maryland 20742, USA
[7] Department of Physics, Applied Physics and Astronomy, Binghamton University, Binghamton, NY 13902
[8] Department of Physics and Astronomy, Rutgers University, Piscataway, New Jersey 08854-8019, USA
[9] Department of Condensed Matter Physics and Materials Science, Brookhaven National Laboratory, Upton, New York 11973, USA

**Supplementary Note 1. Uranium f-electron charge density versus multiplet symmetry**

There are at present very few multiplet-resolving measurements in the literature capable of distinguishing an effective local $5f^1$ symmetry. However, the f-electron charge density can be characterized from a wide range of deeper core levels, and very rarely shows just ~1 f-electron state within uranium compounds. When comparing between multiplet- and density-resolving spectroscopic approaches, one must bear in mind that the lowest order expectation with an electron-rich ligand (like Bi/Sb) and low-f-electron density metal (like U) is that f-electron density should have a significantly higher value than the multiplet symmetry would indicate.

This is well known from a wide range of materials, such as transition metal oxides. For example, numerical models for insulating $VO_2$ give roughly two 3d electrons per vanadium atom (1.83 to 2.48 d-electrons from different modeling approaches in Ref. [1]), even though the multiplet symmetry is unambiguously $3d^1$-like. Likewise, the DFT+DMFT valence histogram restricted to a single-atom basis (as in Fig. 4a of the main text) shows that the occupancy of the $USb_2$ U 5f orbital is 2.17, representing a nominal $5f^2$ valence with weak mixed-valence character due largely to metal ligand hybridization.

**Supplementary Note 2. Exclusion of R2 XAS data due to the Fano effect**

Unlike the case for X-ray measurements, the strong non-resonant photoemission that occurs in the VUV can result in significant Fano interference that greatly modifies broader TEY XAS features [2]. This amplitude of this Fano interference depends on properties of the low energy band structure that are not well captured in a multiplet model. The approximate effect of TEY Fano interference on the XAS spectrum of $URu_2Si_2$ was modeled in our earlier work [3], and the results have been reproduced in Supplementary Figure 1. The simulation shows that Fano interference shifts R2 on a highly problematic energy scale of ~1 eV, while the corresponding energy shift in R1 is quite small (<~0.020 eV).

One might think that measuring XAS by the total fluorescence yield (TFY) method would provide a good workaround. Unfortunately however, extreme ultraviolet TFY tends to be influenced to a similar or greater degree by the elastic photon Fano effect [4], and does not provide an effective route to overcoming the issue.

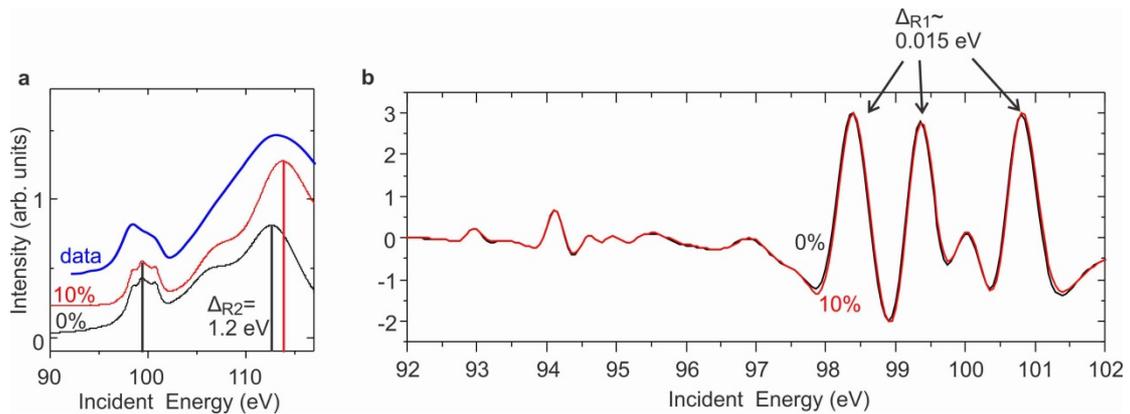

**Supplementary Figure 1: Effect of Fano interference on R2. a,** The O-edge XAS spectrum of (blue) an USb$_2$ surface is compared with (black) a standard multiplet simulation for U$^{4+}$ and (red) the same simulation after incorporating moderate Fano interference. Vertical lines indicate the simulated R1 and R2 intensity maxima, which are offset by $\Delta$ = 1.2 eV by Fano interference at R2. **b**, SDI curves are shown for the 2 simulations, near R1. The primary R1 features are offset by just $\Delta \sim$ 0.015 eV. Panel (a) is adapted from Ref. [4].

**Supplementary Note 3. The uneven surface topography of UBi$_2$.**

The vacuum cleaved surface of USb$_2$ was studied by scanning tunneling microscopy (STM) in Ref. [5], revealing a clean cleave within the [001] plane, and a surface terminated by antimony atoms. No such STM work on UBi$_2$ has previously been reported. An STM topographic map of vacuum-cleaved UBi$_2$ is shown in Supplementary Figure 2, revealing an uneven surface with multiple non-parallel cleavage planes. No large flat area with clear atomic resolution was found for most parts of the sample. Only small, spatially bounded regions were observed with a square lattice structure matching the [001] facet of nominal cleavage (see Supplementary Figure 2a, inset). Rough surfaces of this sort can be highly problematic for techniques specifically sensitive to the outermost atomic layer, such as angle resolved photoemission (ARPES) and STM. The uneven cleavage also limits the quantitative analysis of total electron yield X-ray absorption spectroscopy (TEY XAS), which is expected to have a ~>2 nm probe depth (see Methods).

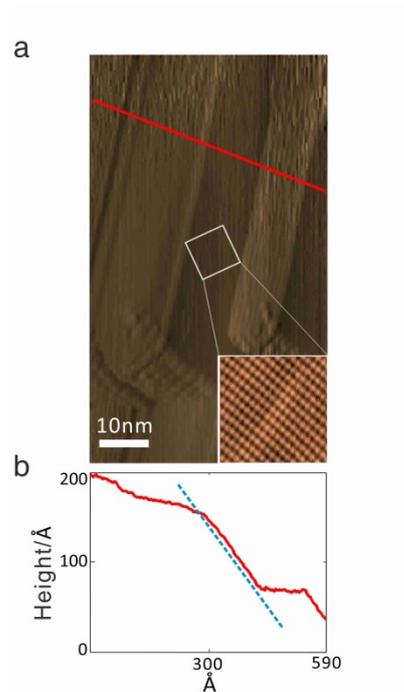

**Supplementary Figure 2: The cleaved surface of $UBi_2$. a,** A 60 *100nm STM topographic map of vacuum-cleaved $UBi_2$ shows multiple non-parallel surfaces. Only select regions can be identified as the [001] plane (insert). The measured in-plane lattice constant is around 0.45nm, matching the lattice constant of $UBi_2$ ($a$=0.4445 nm). **b,** The height profile along the red line in (a). The blue dashed line indicates a terrace with atomic structure matching the (001) facet.

### Supplementary Note 4. The linear dichroic effect in $UBi_2$

The linear dichroic effect at peak-A ($h\nu$=99.2eV) of $UBi_2$ is shown in Supplementary Figure 3a. Other than at the lowest temperature ($T$=15K), the XLD of $UBi_2$ is not significantly distinguished from the systematic error bars from data normalization (~2%). The XLD amplitude is positive throughout the R1 O-edge resonance region (shaded area in Supplementary Figure 3b), and does not show temperature dependence with sufficient amplitude for a close analysis.

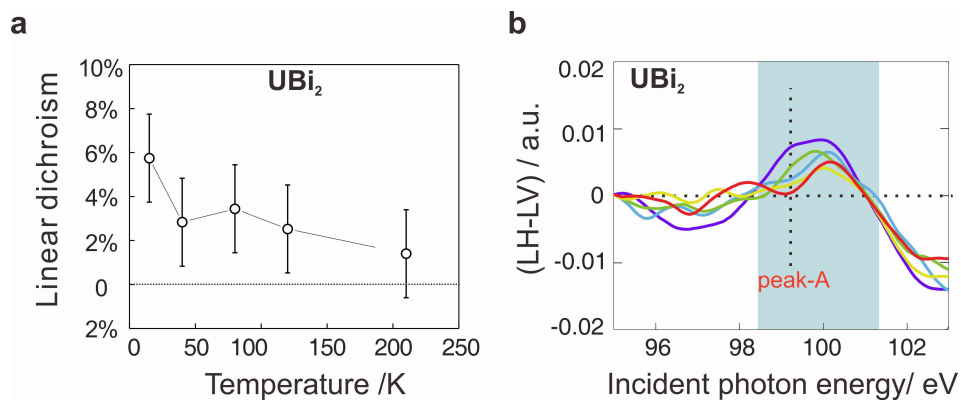

**Supplementary Figure 3: The analysis of XLD of UBi$_2$. a**, The linear dichroism of UBi$_2$. **b**, A shaded area (from 98.5eV to 101.4eV) where the XAS peaks of UBi$_2$ resident. All of this area is dominated by the positive XLD, and lack of negative XLD.

**Supplementary Note 5. CEF symmetries of 5$f^2$ uranium in USb$_2$**

Hund's rule interactions for a single 5$f^2$ uranium atom create a 9-fold degenerate ground state basis with angular momentum $J=4$. These states are further split by the crystal electric field (CEF) into five singlet states and two doublets. The singlet states include $\Gamma_1$ ($|m_J|=0$, 4 moment components), $\Gamma_2$ ($|m_J|=4$) and $\Gamma_3$ ($|m_J|=2$). The doublets are derived from $|m_J|=1$, 3 components, and include the $\Gamma_5$ state with occupancy comparable to the CEF ground state at T>~100K in the AM+MF model. These symmetry labels are identical to those of URu$_2$Si$_2$ [4, 6], and represent mixing over +/-4 units of angular momentum by a crystal field with $C_4$ rotational symmetry.

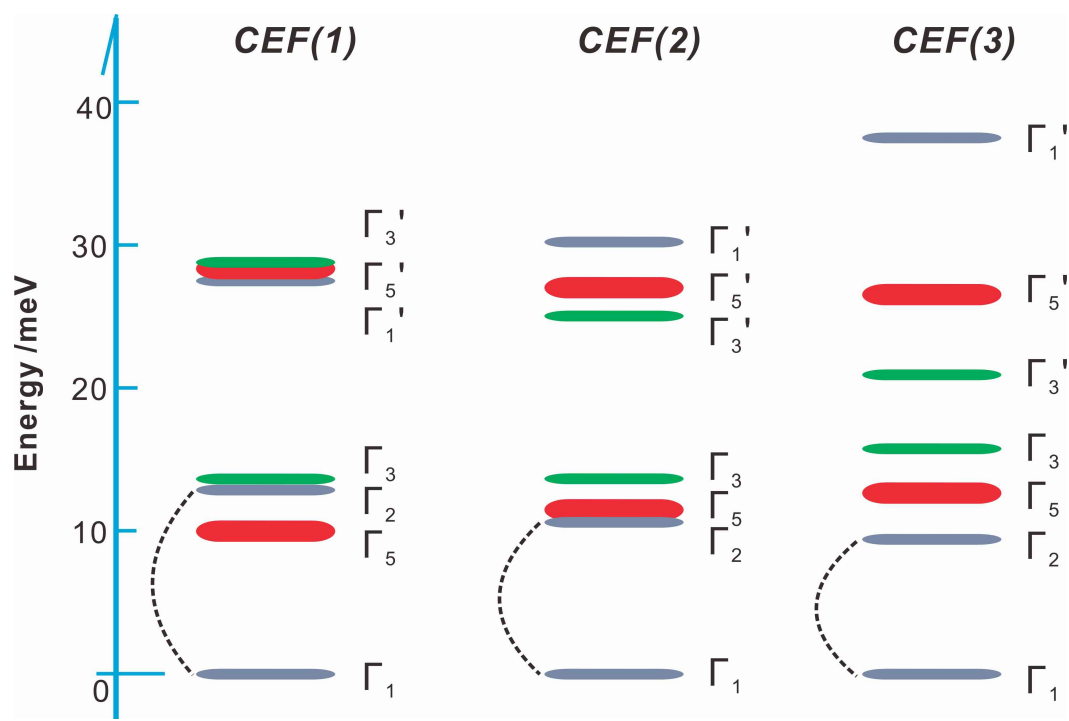

**Supplementary Figure 4: The paramagnetic CEF symmetries of 5$f^2$ uranium.** The *CEF(1)* crystal field scheme is compared with alternatives (*CEF(2)* and *CEF(3)*), as defined by paramters listed in table 1 of the main text. The doublet states ($\Gamma_5$ and $\Gamma_5^1$) are plotted with thicker lines. A dashed line is drawn connecting the $\Gamma_1$ and $\Gamma_2$ singlets, which can generate z-axis magnetic moment when combined coherently. Beneath the Neel temperature, the $\Gamma_1$ state becomes mixed with $\Gamma_2$, and the $\Gamma_5$ doublet splits into magnetically polarized $\Gamma_5(1)$ and $\Gamma_5(2)$ branches

**Supplementary Note 6. Oxygen $L_1$-edge XAS**

The $USb_2$ and $UBi_2$ samples are surveyed by XAS at the oxygen $L_1$-edge in Supplementary Figure 5. Both samples are found to have a broad peak around 42eV, which is very similar to the oxygen $L_1$-edge seen from a partially oxidized iron selenide (O-FeSe) sample (Supplementary Figure 5a). Negative second derivative curves show two distinctive sub-peaks (Supplementary Figure 5b). The first peak is located at 41.6eV (peak A), and is found in all three samples. The second one is at 43eV (peak B), within ~1eV of the nominal binding energy of uranium 6s-electrons (uranium $P_1$-edge), and is absent for the FeSe sample. By fitting the $USb_2$ and $UBi_2$ spectra with Voigt functions at peak A and peak B, we find that the ratio of peak intensities *I*(A)/*I*(B) is 15% smaller for $USb_2$, suggesting that there may be approximately 15% less oxygen at the $USb_2$ surface if the 43eV peak is attributed to uranium. This 15% value is comparable to error bars from the background subtraction and fitting procedures, and one can only conclude that the measurement is suggestive of a similar oxygen concentration at the surface of both samples.

The above analysis is complicated by the uncertainty in the attribution of peak B. It is worth noting that the absolute intensity of the ~41eV XAS signal was also comparable between the two samples, with somewhat more intensity (47% more) fitted in the $UBi_2$ spectrum. This corroborates the idea that there may be a higher oxygen concentration at the surface of this sample, but does nothing to improve error bars, as absolute XAS intensity tends to be unreliable when comparing between different samples.

The role of oxygen at the surface is unclear, but one can assume that some is present in the form of adsorbed polar molecules such as $H_2O$, CO, and $CO_2$, which are present in the vacuum, and will play a role in screening the polar surface presented by the $UX_2$ structure.

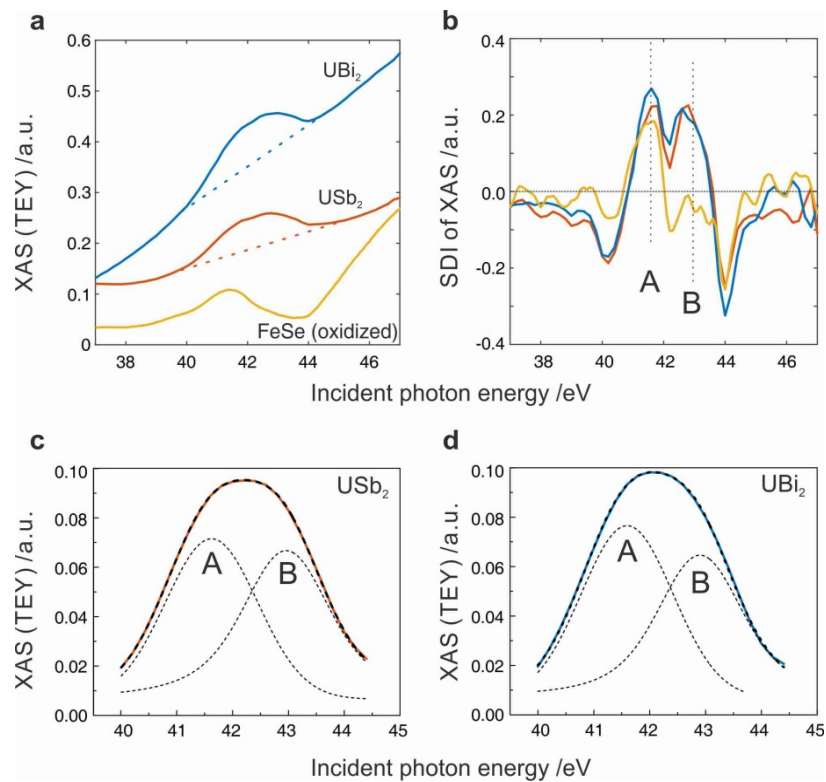

**Supplementary Figure 5: Oxygen L-edge XAS. a,** The XAS (TEY) spectra of $UBi_2$ (blue), $USb_2$ (red), and a partially oxidized FeSe sample (yellow). **b,** The negative second derivative curves of XAS spectra in (a) Two peaks are labeled A (41.6eV) and B (43.0eV). **c-d,** The XAS of $USb_2$ (c) and $UBi_2$ (d) are fitted with two Voigt functions after subtraction of the linear backgrounds drawn in panel (a). The raw data in panel (c) are scaled up by a relative factor of 1.5, compared with raw data underlying panel (d).

**Supplementary Note 7. Hartree-Fock renormalization**

In the main text, two different sets of Hartree-Fock parameters are used to simulate the XAS spectra of $UBi_2$ and $USb_2$, as described in Methods. Supplementary Figure 6 shows that the XAS simulations remain very similar if these Hartree-Fock parameter sets are reversed.

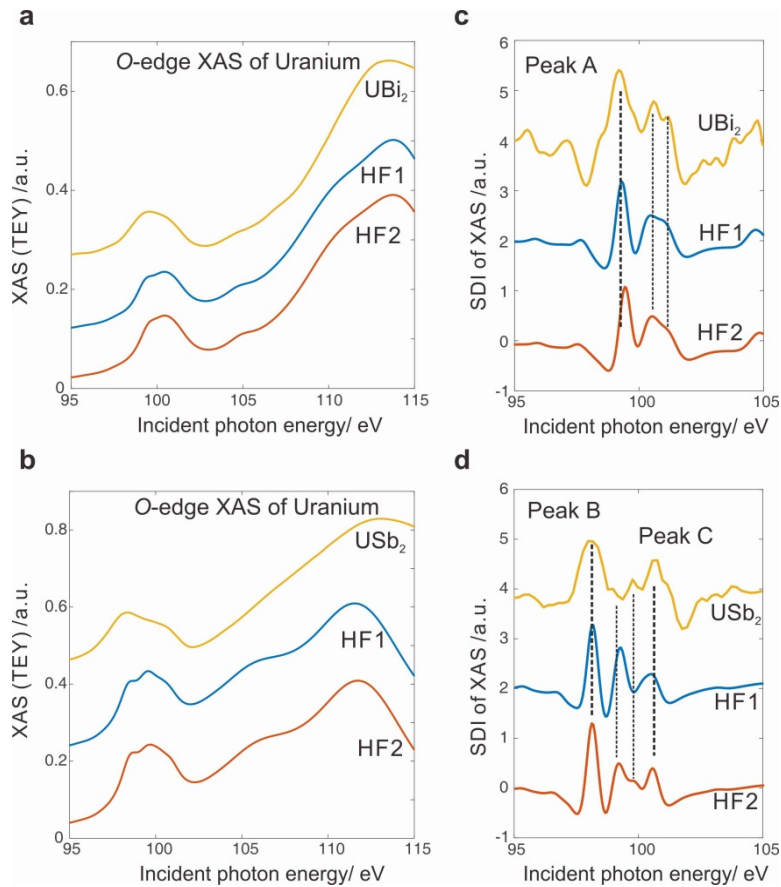

**Supplementary Figure 6**: **Hartree-Fock renormalization and the XAS spectrum. a-b,** The uranium O-edge XAS spectra of (a) $UBi_2$ and (b) $USb_2$ are compared with simulated XAS spectra fitted with the parameter sets used for $UBi_2$ (labeled HF1) and $USb_2$ (labeled HF2). **c-d,** The negative SDI curves extract from the experimental XAS data and simulated curves in (a) and (c) The labels Peak-A (98.2eV), Peak B (99.2eV) and Peak C (100.8eV) are used to mark the main features in the XAS spectra, as in the main text.

## Supplementary References

*Philosophical Magazine*, **89**:22-24, 1881-1891 (2009).

[6] Sundermann, M. et al. Direct bulk sensitive probe of 5f symmetry in $URu_2Si_2$, *Proc. Natl. Acad. Sci. (USA)* **113**, 13989 (2016).